# Multiple Tunable Hyperbolic Resonances in Broadband Infrared Carbon-Nanotube Metamaterials

*John Andris Roberts[1,‡], Po-Hsun Ho[2,‡], Shang-Jie Yu[2,‡], Xiangjin Wu[2], Yue Luo[3], William L. Wilson[3], Abram L. Falk[4]\*, and Jonathan A. Fan[2]\**

[1]Department of Applied Physics, Stanford University, Stanford, CA 94305, U.S.A.

[2]Department of Electrical Engineering, Stanford University, Stanford, CA 94305, U.S.A.

[3]Center for Nanoscale Systems, Harvard University, Cambridge, MA 02138, U.S.A.

[4]IBM T.J. Watson Research Center, Yorktown Heights, NY 10598, U.S.A.

‡ These authors contributed equally to this work.

*To whom correspondence should be addressed (alfalk@us.ibm.com, jonfan@stanford.edu)

**Abstract**

Aligned, densely-packed carbon nanotube metamaterials prepared using vacuum filtration are an emerging infrared nanophotonic material. We report multiple hyperbolic plasmon resonances, together spanning the mid-infrared, in individual resonators made from aligned and densely-packed carbon nanotubes. In the first near-field scanning optical microscopy (NSOM) imaging study of nanotube metamaterial resonators, we observe distinct deeply-subwavelength field profiles at the fundamental and higher-order resonant frequencies. The wafer-scale area of the nanotube metamaterials allows us to combine this near-field imaging with a systematic far-field spectroscopic study of the scaling properties of many resonator arrays. Thorough theoretical modeling agrees with these measurements and identifies the resonances as higher-order Fabry-Pérot (FP) resonances of hyperbolic waveguide modes. Nanotube resonator arrays show broadband extinction from 1.5-10 μm and reversibly switchable extinction in the 3-5 μm atmospheric transparency window through the coexistence of multiple modes in individual ribbons. Broadband carbon nanotube metamaterials supporting multiple resonant modes are a promising candidate for ultracompact absorbers, tunable thermal emitters, and broadband sensors in the mid-infrared.

**Main Text**

Recently, large-area, densely packed, and globally aligned films of carbon nanotubes prepared using solution-based vacuum filtration have emerged as a promising material for nanophotonics [1,2]. Nanotube films have demonstrated capabilities including strong and reversible tunability via doping [3], polarization-sensitive light detection and thermal emission [1,4,5], nonlinear optics [6,7], and ultrastrong exciton-cavity coupling [8–10]. Moreover, aligned nanotube films function as optical hyperbolic metamaterials across the mid-infrared, with an effective dielectric permittivity that is negative along the nanotube alignment axis and positive in the perpendicular directions from 1.7 μm - 6.7 μm wavelengths for highly doped films [11,12]. Like other hyperbolic materials [13–19], aligned nanotubes can support highly confined hyperbolic plasmons in nanostructures [11,20–23]. This combination of functional capabilities and hyperbolic dispersion makes aligned nanotubes a promising nanophotonic platform for emission, detection, and polarization control of light. The solution-based preparation of the films also offers unique benefits like the ability to assemble nanotubes of purified chirality [9,24] and to transfer assembled large-area films with controlled thicknesses to arbitrary substrates.

An open question is whether aligned nanotube metamaterials can support multiple hyperbolic resonances. Multiple resonances at different frequencies, including higher-order modes, have been observed in metallic nanostructures [25–29]. Because of their broadband hyperbolic dispersion, carbon nanotube metamaterials offer to provide a similar capability in the mid-infrared. In this wavelength range, they have the potential to combine the active tunability and deeply subwavelength confinement achieved in graphene plasmonics [30–33] with the enhanced coupling to free-space radiation enabled by dense nanotube packing in a thick film [21]. Multiple hyperbolic resonances across this wavelength range could be combined with the unique functional capabilities of nanotubes for new applications. The temperature stability and low thermal mass of carbon nanotubes could be combined with resonant spectral selectivity at multiple frequencies to enable thermal emitters with tailored spectral properties [20,34–38]. Nanotubes supporting multiple resonances could be used for multiband surface-enhanced infrared absorption (SEIRA) with actively tunable resonant frequencies [39,40]. Finally, carbon nanotube resonators with broadband extinction enabled by multiple modes could serve as compact, low-thermal mass absorbers for photodetection [41–43].

In this work, we show that individual resonators patterned from thick films of aligned carbon nanotubes can support several resonant modes that together span the entire mid-infrared. We perform near-field imaging to visualize the field profiles of these modes, correlating with far-field spectra in order to select resonant frequencies. To understand the origin of the resonances, we use the large area of the nanotube films to study many arrays of resonators with varying parameters. We conclude that the resonances are hyperbolic waveguide Fabry-Pérot (FP) modes. With the strong far-field coupling enabled by dense nanotube packing in a thick film, we demonstrate nanotube resonators with broadband extinction from the longwave infrared to the near-infrared that can be reversibly switched through doping.

When relatively thick (tens to hundreds of nanometers) films of aligned carbon nanotubes are patterned into ribbons, extinction peaks appear in the films' mid-infrared spectra for incident light polarized along the nanotubes [8,21]. These extinction peaks are the result of the fundamental FP resonances of hyperbolic waveguide modes in the ribbons [11]. Because of their hyperbolic dispersion, nanotube films support in-plane waveguide modes with fields tightly confined to the film and decaying into the air and substrate. In a nanotube film supporting a hyperbolic guided wave, a patterned ribbon will be resonant when the FP condition $2\beta L+2\phi=2\pi m$ is satisfied [Fig. 1(a)], where $\beta$ is the wavevector of the hyperbolic waveguide mode at the resonant frequency, $\phi$ is the phase picked up by the hyperbolic waveguide mode on reflection from the ribbon edge, and *m* is an integer. Wider ribbons are resonant at lower frequencies where the guided waves have longer wavelengths. This model has previously been applied to understand the resonances of metallic nanoantennas [28,44]. In long metallic nanoantennas and nanowires, multiband resonant phenomena across wide wavelength ranges are enabled by plasmonic standing-wave FP resonances that are well-separated in frequency [25,29,45]. In carbon nanotube metamaterials, only the *m* = 1 mode has been identified, in ribbons with widths of several hundred nanometers [11].

We report the appearance of higher-order hyperbolic resonances in microscale aligned carbon nanotube ribbons [Fig. 1(b)]. We prepare films of aligned, unsorted single-walled carbon nanotubes (SWCNTs) using vacuum filtration and transfer them to silicon substrates with a thin $HfO_2$ layer. We pattern the films into arrays of ribbon resonators with varying widths using electron-beam lithography and oxygen plasma etching, and reversibly dope the films by exposing them to nitric acid vapor. With this doping, we expect the films to be optically hyperbolic in the mid-infrared, with a negative permittivity along the nanotube direction ($\varepsilon_z < 0$) and a positive permittivity perpendicular to the nanotubes ($\varepsilon_{xy} > 0$) (see Fig. S10) [12]. We measure the extinction spectra of the ribbon arrays using Fourier-transform infrared (FTIR) spectroscopy. In highly-doped ribbons with widths less than 600 nm, there is one prominent peak in the extinction spectra which redshifts with increasing ribbon width. This peak corresponds to the *m* = 1 FP resonance of a hyperbolic waveguide mode [11].

For ribbons with widths greater than 560 nm, a second resonance appears in the extinction spectra at higher frequencies (Fig. 1(b), second dashed line), and for ribbons wider than 1.5 μm a third peak appears (Fig. 1(b), third dashed line. Also see Figs. S3 and S5). These extinction peaks also redshift as the ribbon width increases. To theoretically investigate the nature of these extinction peaks, we perform finite-difference time domain (FDTD) simulations of the ribbons, treating them as bulk materials with an anisotropic effective dielectric tensor previously extracted using spectroscopic ellipsometry [11,12]. We obtain simulated extinction spectra that are in general agreement with our experimental results [Fig. 1(c)], although there are discrepancies that may be due to higher levels of doping in the resonators than in the samples studied by ellipsometry (see the Supplemental Material). Notably, the widest ribbons have fundamental extinction peaks at frequencies below the hyperbolic range in the simulations, where the nanotube films are similar to an anisotropic metal with $\varepsilon_z < 0$ and $\varepsilon_{xy} < 0$. Based on our FP resonator model, and confirmed by simulated field profiles, near-field imaging, and guided-mode analysis as discussed below, we identify the higher-frequency peaks as the $m = 3$ and $m = 5$ FP resonances of hyperbolic waveguide modes. The even-numbered resonances do not appear in the extinction spectra because they have symmetric field profiles and therefore do not couple to free-space radiation [28,46].

One key signature of the nanoresonator modes is their local field patterns, which uniquely correspond to modes of different orders. However, these local fields cannot be resolved by far-field IR microscopy because the optical features of the highly confined modes are at a deeply-subwavelength scale. Near-field scanning optical microscopy (NSOM) couples free-space light into the near field via a sharp probe. It can readily reach a spatial resolution of ~20 nm and has been employed to study various types of deep subwavelength polaritons [47–49]. In our NSOM experiments, we image individual ribbon resonators at their fundamental and higher-order resonant frequencies [Fig. 2(a)]. We use a Neaspec NSOM system equipped with tunable mid-infrared lasers for single-frequency imaging. We choose 1 µm- and 2.5 µm-wide ribbon arrays that clearly support $m = 1$ and $m = 3$ resonances in the far-field FTIR spectra of the sample under the moderately doped condition [Fig. 2(b)]. The NSOM probing wavelengths we select correspond to the $m = 1$ resonance of the 1 µm ribbon ($\lambda = 10.6$ µm) and the $m = 3$ resonance of the 2.5 µm ribbon ($\lambda = 6.6$ µm). Before the NSOM measurement, we chemically dope the sample to a moderate doping level by exposing it to nitric acid vapor (see the Supplemental Material for detailed doping procedure).

At these frequencies, we obtain the raw NSOM images shown in Fig. 2(b). The near-field profiles are clearly asymmetric in the direction perpendicular to the ribbon, which is a sign of the presence of a localized mode. When we image the near-field pattern in the nanostructures with NSOM, there are two main contributions to the complex amplitude maps [50], which are associated with different pathways to excite and outcouple the modes. The first contribution is the round-trip contribution and involves both plasmonic excitation and outcoupling through the NSOM tip scattering [Fig. 2(a), bottom left]. The round-trip contribution has a symmetric near-field profile because of the structural symmetry. The second contribution is direct coupling to the localized mode, which can be either excited by the tip and scattered back at the nanostructure edge, or excited at the edge and picked up by the tip [Fig. 2(a), bottom right]. The direct coupling contribution has an antisymmetric near-field profile because of its dipolar nature. The localized modes we have seen with far-field spectroscopy can be nanoimaged by extracting the direct coupling contribution of the NSOM field map. The total signal is asymmetric since there are both round-trip (symmetric) and direct coupling (antisymmetric) contributions. This measurement is done under a perpendicular configuration (the incident $k$-vector is perpendicular to the nanotube ribbon cut direction) since this way can effectively excite the direct coupling contribution (see the Supplemental Material for the detailed discussion on the incident *k*-vector direction) [51].

We average the complex NSOM signal (amplitude and phase) over the length of the ribbon and take the antisymmetric part to obtain the direct coupling component resulting from the localized mode [Fig. 2(c)]. For comparison, we simulate the field profiles of the modes in the ribbon. The NSOM tip is sensitive to the electric field component normal to the surface, $E_y$. We plot the simulated E-field cross-section Re($E_y$) and find a good qualitative match between the NSOM data and the simulation. The sharp peaks on the two sides in both signals are the local hot spots at the ribbon edges. For a comparison to the expected field profile of a FP resonator, we can examine the number of field nodes inside the resonator. For the 1 μm ribbon at the fundamental frequency, there is one node in the center of the ribbon. The signal from the 2.5 μm ribbon at the higher-order resonant frequency shows three nodes near the center of the ribbon, a sign that the ribbon supports an *m* = 3 FP mode at this frequency. Although, in simulation, this field pattern is localized inside the ribbon, the experimental data shows a similar tendency because the NSOM probe can access the subsurface fields [52,53].

Having identified the field profiles of the modes, we show that they are higher-order FP resonances of hyperbolic guided waves through a systematic study of their scaling properties. This study is enabled by the large area, global alignment, and thickness uniformity of the nanotube films. Across a single nanotube film, we pattern many resonator arrays, varying not only the ribbon width *L* but also the cut angle *θ* of the ribbon relative to the nanotube alignment axis [Fig. 3(a)]. We observe that when the cut angle of the ribbon varies away from normal to the nanotube alignment axis, so that the nanotubes are diagonal within the ribbon, the higher-order extinction peak redshifts for a fixed ribbon width [Fig. 3(b), second dashed line]. We have previously identified this redshift for the fundamental hyperbolic mode [11]. The redshift appears because, for a diagonal cut angle, the resonant guided mode wavevector approaches a high-$k$ tangent of the in-plane hyperbolic isofrequency surface [Fig. 3(a)]. This makes it possible to confine light in a deeply subwavelength volume. Equivalently, ribbons of a fixed width become resonant at lower frequencies where they are smaller compared to the free-space wavelength. The fundamental resonance in these ribbons can be at a frequency lower than the hyperbolic range, where the carbon nanotubes are similar to an anisotropic metal (see Fig. S10), and does not show such a dramatic redshift [Fig. 3(b), first dashed line]. The observation of a significant redshift for the higher-order resonance establishes that it is a hyperbolic resonance and validates a prediction of the effective-medium hyperbolic metamaterial picture.

Using data from many highly doped resonator arrays with different widths and cut angles, we construct the dispersion relations of the hyperbolic waveguide modes and conclusively demonstrate that the extinction peaks we observe are the $m = 1, 3$, and 5 FP resonances of these modes. For a ribbon with width $L$, the wavenumber of the resonant hyperbolic guided wave is $\beta = (m\pi - \phi)/L$. Knowing $L$, estimating $\phi \approx \pi/4$ based on simulations (see Supplemental Material), and hypothesizing that the resonances correspond to $m = 1, 3$, and 5, we extract the guided wavenumber $\beta$ from each resonance. The resonant frequency is identified by peak fitting (see Fig. S5). With this information, we construct the wavenumber-versus-frequency dispersion of the guided waves for different propagation directions (corresponding to different cut angles $\theta$) using data from many ribbons. With these hypothesized mode indices, the dispersion relations fall on a single line for each direction [Fig. 3(c)]. This demonstrates that we have correctly identified the waveguide mode origin and FP mode number. At smaller cut angles, the dispersion is redshifted as the guided mode wavevector approaches the high-$k$ tangent of the hyperbolic isofrequency surface, in agreement with our previous theoretical analysis [11]. The observation of hyperbolic dispersion also further validates the effective medium model of the nanotube metamaterial, demonstrating that the effective dielectric functions can be used to predict and design complex resonant behavior in the same way that bulk dielectric functions are used for metals.

The coexistence of multiple resonances enables broadband extinction from the far-infrared to the mid-infrared. To maximize the extinction across a broad wavelength range, we use a thicker film (195 nm) that supports higher-frequency resonances and more uniform broadband extinction (see Figs. S7 and S13). In the undoped state, the extinction spectrum of a resonator array with ribbon width 1.5 µm and thickness 195 nm for polarization along the nanotubes shows fundamental and higher-order peaks around $\lambda$ = 10 µm and a peak in the near-infrared near $\lambda$ = 2 µm caused by the $S_{11}$ exciton in the semiconducting nanotubes [Fig. 4(a)] [54]. Between these peaks, there is low extinction in the mid-wave infrared (MWIR) from $\lambda$ = 3 - 5 µm. Immediately after doping by exposure to nitric acid vapor, the exciton peak is quenched and three distinct resonant peaks cause extinction from $\lambda$ = 2.5 - 10 µm. At moderate and ambient doping levels, the extinction becomes less broadband, and at ambient doping levels the $S_{11}$ exciton peak begins to reappear (see Supplemental Material for definitions and characterization of doping levels). When the vapor doping is removed by annealing, the film reverts to its original state, with low extinction in the MWIR. In this way, we are able to reversibly switch the extinction in the technologically important 3 – 5 µm range. The dense packing of the thick nanotube films enables overall large extinction, while retaining the tunability of systems such as thin carbon nanotube films and graphene.

Simulations of the resonator array using the dielectric functions for moderate doping [Fig. 4(b)] allow us to identify these resonances as the $m$ = 1, 3, and 5 FP resonances. The ribbons' ability to support these resonances simultaneously enables the broadband extinction. The lowest-frequency resonance occurs in the simulation at frequencies lower than the hyperbolic range, where the film is still anisotropic but both permittivies are negative. This resonance, in the anisotropic metal-like frequency range, is similar to the fundamental peak in Fig. 3(c). Experimentally, the frequency of this peak shifts the least with doping as shown in Fig. 4(a). The existence of an anisotropic-metal-like $m = 1$ resonance outside the hyperbolic range provides a mechanism to complement the hyperbolic resonances and provide very broadband extinction. By studying the simulated field profiles of the other two resonances [Fig. 4(b)], we identify the second and third extinction peaks as the $m = 3$ and $m = 5$ hyperbolic FP resonances. This identification is confirmed by the scaling properties of the resonances [Fig. S4]. The combination of resonances with distinct plasmonic and hyperbolic origins enables broadband extinction spanning the MWIR as well as longer wavelengths and approaching the near-infrared, and demonstrates the variety of phenomena occurring in the nanotube metamaterial as well as the advantage of using wide ribbons to support multiple resonances.

In summary, we have demonstrated that carbon nanotube metamaterials can simultaneously support multiple resonances. We observe the appearance of higher-order modes in microscale nanotube ribbon resonators using far-field infrared spectroscopy and visualize them by extracting their field profiles from infrared near-field images. This combination of far-field spectroscopy and near-field visualization is a powerful technique to study localized modes in resonators. The wafer-scale area, thickness uniformity, and global alignment of the nanotube films allows us to perform a systematic study of the resonators' scaling properties. This study shows that the resonances are the $m = 1, 3,$ and 5 FP resonances of hyperbolic guided waves in the nanotube films. We compare these observations to predictions based on effective medium optical constants and find good agreement, showing that the metamaterial picture of aligned nanotube films can be used to predict and design complex resonant behavior. We demonstrate a ribbon array with multiple resonances that supports broadband extinction in the infrared and can be reversibly switched in the technologically important MWIR region through doping control.

Our work establishes that aligned, densely-packed carbon nanotube metamaterials can support multiple resonant modes, a pathway for applications requiring tunable broadband or multiband performance in the mid-infrared. Higher-order modes have long been observed in noble metal plasmonic resonators, and our work places aligned nanotube films on a trajectory to provide similar capabilities in the mid-infrared. We expect that advances in processing [55,56] will steadily improve the performance of aligned nanotube metamaterials, potentially toward the limit of high-quality-factor plasmons observed in single nanotubes [57]. Aligned nanotubes are potentially actively tunable through ionic liquid gating or electrostatic gating [3] combining the strengths of graphene plasmonics with strong coupling to far-field radiation enabled by dense packing. This combination will make aligned nanotube films candidates for applications such as tunable multiband SEIRA, spectrally engineered thermal emission, and broadband absorption for photodetection. With the ease of large-area fabrication allowed by solution processing, nanotube metamaterial resonators can be the foundation for nanophotonic devices with exciting new capabilities.

**Figures**

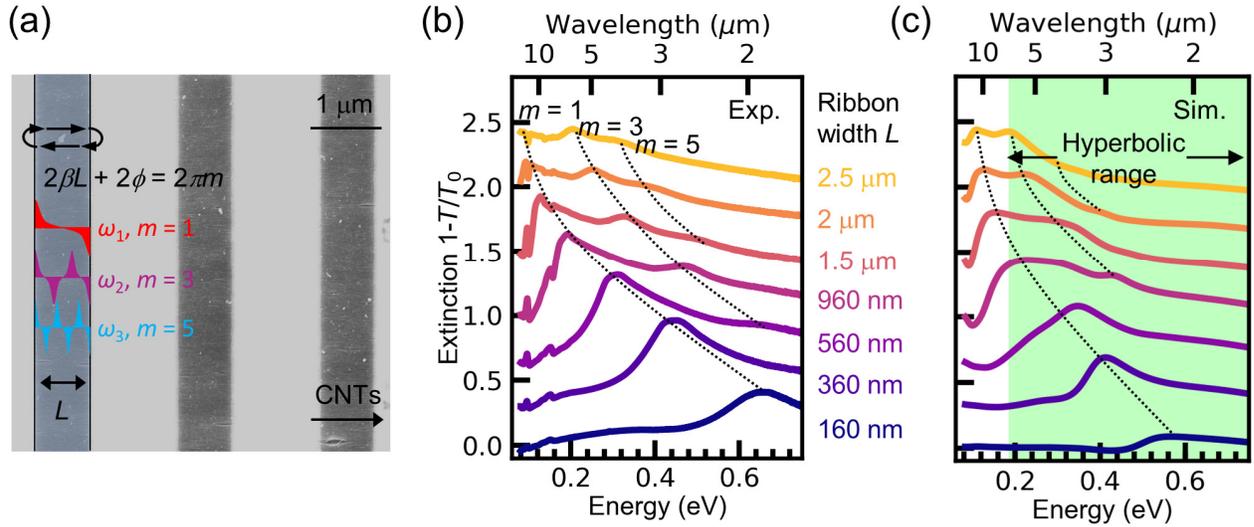

Fig. 1. (a) SEM image of ribbon resonators patterned in a thick (approximately 80 nm) film of aligned, unsorted single-walled carbon nanotubes, overlaid with a schematic depicting the FP resonant condition for hyperbolic waveguide modes and cartoons of the field profiles for FP resonances of different orders. (b) Experimental far-field extinction spectra of highly doped nanotube resonator arrays patterned in the film shown in (a) with varying ribbon widths, measured using FTIR spectroscopy immediately after doping with nitric acid vapor. In wider ribbons, higher-order resonances appear. The dashed lines are guides to the eye. Spectra are offset vertically for clarity. (c) Simulated extinction spectra corresponding to the measurement in (b) using effective dielectric functions for a moderately doped nanotube film in a FDTD model. The green shaded area represents the wavelength range where the moderately doped nanotube film has hyperbolic dispersion.

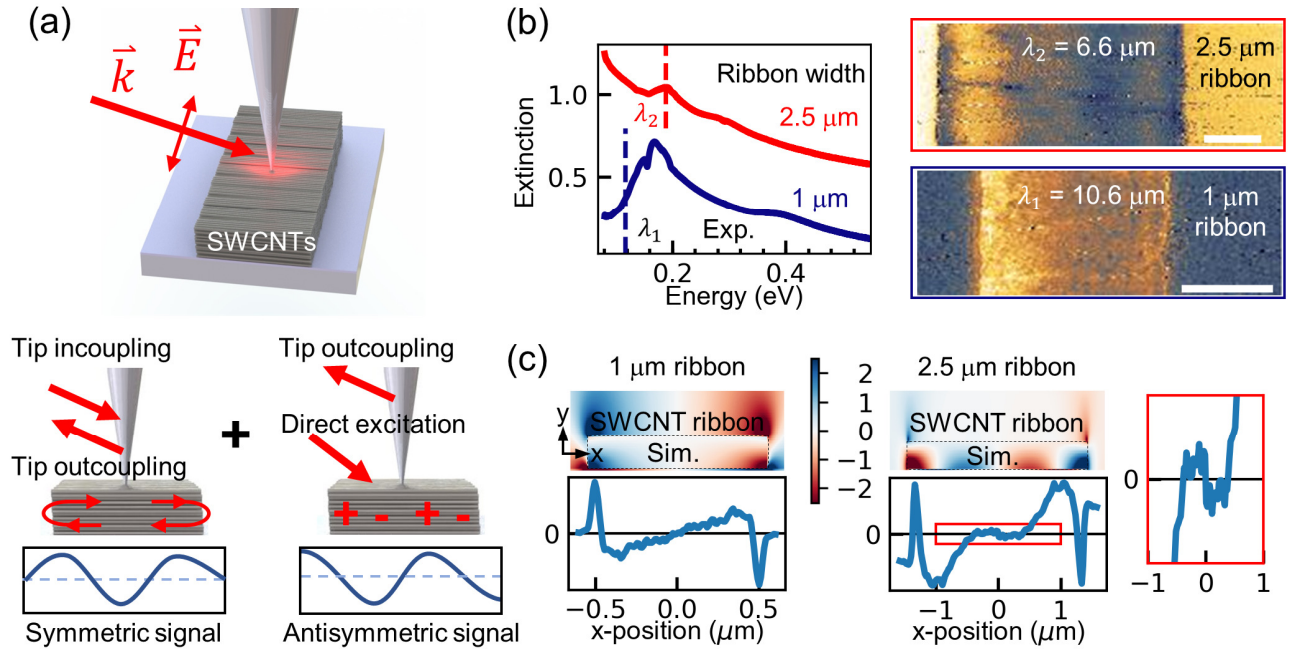

Fig. 2. (a) Cartoons of NSOM measurement of resonant modes in a nanotube ribbon resonator, showing the incident light with wavevector $k$ perpendicular to the ribbon and the tip used to measure the near-field signal (top) and cartoons showing the symmetric round-trip and antisymmetric direct-coupling signals. (b) Far-field extinction spectra of two resonator arrays with ribbon widths 1 μm and 2.5 μm and thickness 85 nm (vertically offset for clarity), showing the wavelengths used for the near-field imaging experiment. For the 2.5 μm ribbon, the laser is chosen to coincide with the higher-order peak, and for the 1 μm ribbon the laser wavelength is chosen to measure only the fundamental resonance. Right, NSOM phase images of the two ribbons at the chosen wavelengths. The asymmetry is a sign of the presence of a localized resonance. Scale bars: 500 nm. (c) Real component of the antisymmetric part of the NSOM signals (experiment), corresponding to the signal from the localized modes, compared to the simulated field $Re(E_y)$ of the ribbon resonators, showing qualitative agreement. The center part of the NSOM signal for the 2.5 μm ribbon is shown in more detail in the red-outlined inset.

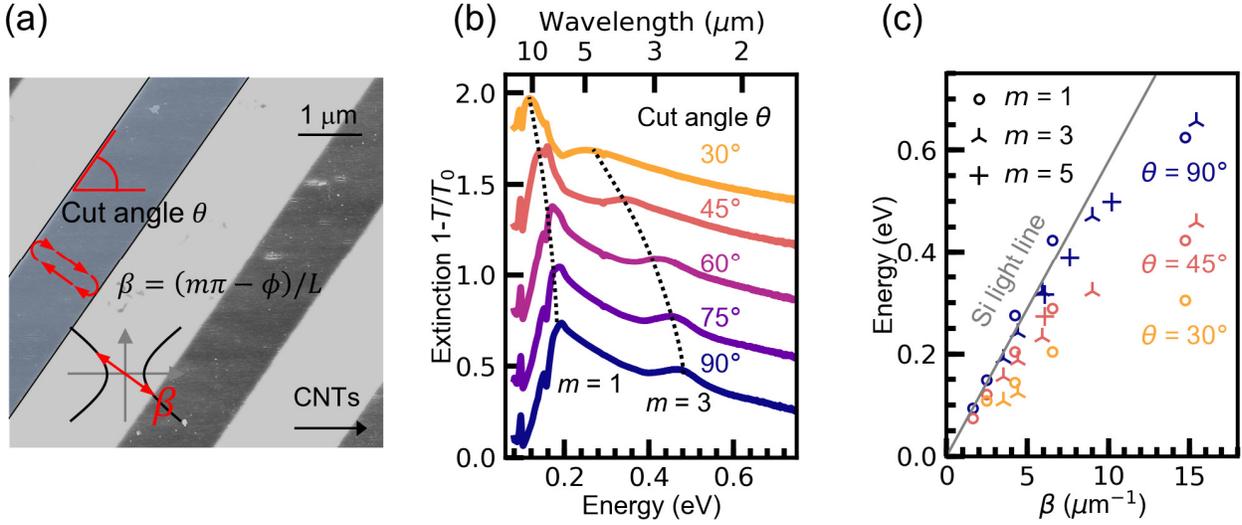

Fig. 3. (a) SEM image of a nanotube film (thickness approximately 80 nm) overlaid with a cartoon showing the cut angle $\theta$ between the ribbon direction and the nanotube alignment and indicating the direction of the resonant wavevector and its magnitude. The bottom cartoon shows the resonant wavevector on the nanotube hyperbolic isofrequency surface. (b) Experimental FTIR extinction spectra of ribbon arrays with width 1 μm, thickness approximately 80 nm, and a range of cut angles (vertical offset for clarity). As the cut angle is decreased, the $m = 3$ peak redshifts because the resonant wavevector samples high-$k$ points on the hyperbolic isofrequency surface. (c) Experimental dispersion plot of the hyperbolic waveguide modes constructed using the FP relation and the extinction spectra of many resonator arrays. When the mode orders $m = 1$, 3, and 5 are used for the fundamental and higher-order peaks, the dispersion relation for a given direction follows a single line, confirming that these are the correct FP mode orders. For smaller cut angles, the hyperbolic waveguide mode becomes more highly confined as expected from the in-plane hyperbolic dispersion. The reflection phase is estimated as $\phi \approx \pi/4$ based on simulations.

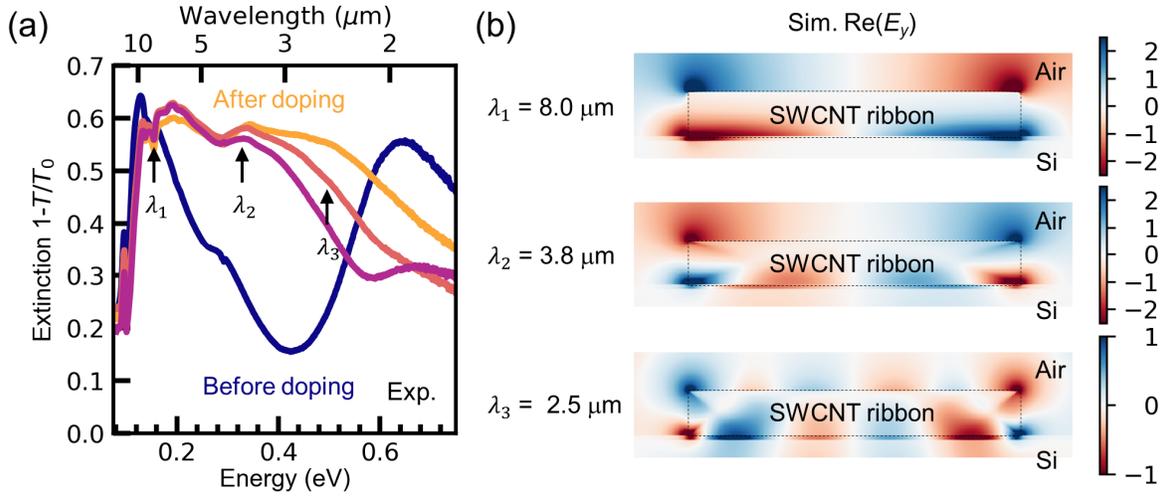

Fig. 4. (a) FTIR extinction spectra of a nanotube resonator array with ribbon width 1.5 μm and thickness approximately 195 nm before (blue) and immediately after (yellow) vapor doping for incident light polarized along the nanotubes, showing broadband reversible extinction enabled by multiple resonances. Spectra at moderate (one day after vapor doping) and ambient doping (five days after vapor doping) are shown in orange and purple, respectively. Wavelengths of the three resonances ($\lambda_1$, $\lambda_2$, and $\lambda_3$) are marked with arrows. (b) Simulated field profiles at the three wavelengths labeled in (a) in a moderately doped ribbon with width 1.5 μm and thickness 200 nm. Top: an $m = 1$ resonance in the region where the nanotube films are similar to an anisotropic metal; middle: the $m = 3$ hyperbolic resonance; bottom: the $m = 5$ hyperbolic resonance.

ACKNOWLEDGMENT

This work was supported by the Air Force Office of Scientific Research (AFOSR) Multidisciplinary University Research Initiative (MURI) under Award FA9550-16-1-0031. J.A.R. was supported by the Department of Defense through the National Defense Science and Engineering Graduate Fellowship Program. Part of this work was performed at the Stanford Nano Shared Facilities (SNSF), supported by the National Science Foundation under award ECCS-1542152 and the Harvard Center for Nanoscale Systems (CNS) supported by the NSF under award NNCI-1541959. Y.L. was supported by the Department of Energy under award DE-SC0019300.